\documentclass[%
nofootinbib,
 reprint,
 amsmath,amssymb,
 aps,
]{revtex4-1}

\usepackage{url,comment}
\usepackage{times}
\usepackage{latexsym}
\usepackage{graphicx, graphics, hyperref, amsmath, amssymb, slashed, xcolor, bbm,amsthm, array,textcomp,gensymb}
 \usepackage{subfigure}

\usepackage{rotating}
\usepackage{afterpage}

\newcommand{\nc}{\newcommand}
\nc{\beq}{\begin{equation}}
\nc{\eeq}{\end{equation}}
\nc{\barray}{\begin{eqnarray}}
\nc{\earray}{\end{eqnarray}}
\nc{\barrayn}{\begin{eqnarray*}}
\nc{\earrayn}{\end{eqnarray*}}
\nc{\bcenter}{\begin{center}}
\nc{\ecenter}{\end{center}}
\nc{\mc}{\mathcal}
\nc{\er}[1]{(\ref{eq:#1})}
\nc{\onehalf}{\frac{1}{2}} 
\nc{\partialbar}{\bar{\partial}}
\nc{\psit}{\widetilde{\psi}}
\nc{\Tr}{\mbox{Tr}}
\nc{\hc}{\mbox{H.c.}}
\nc{\ev}{\;\mathrm{eV}}
\nc{\mev}{\;\mathrm{MeV}}
\nc{\gev}{\;\mathrm{GeV}}
\nc{\tev}{\;\mathrm{TeV}}

\def\chii0{\chi_i^0}
\def\chij0{\chi_j^0}

\newcommand{\gsim}{\lower.7ex\hbox{$\;\stackrel{\textstyle>}{\sim}\;$}}
\newcommand{\lsim}{\lower.7ex\hbox{$\;\stackrel{\textstyle<}{\sim}\;$}}
\nc{\ttbar}{t\bar t}

\def\iab{{\ \rm ab}^{-1}}

\newcommand{\fref}[1]{Fig.~\ref{f.#1}}
\newcommand{\eref}[1]{Eq.~(\ref{e.#1})}

\newcommand{\cref}[1]{Chapter~\ref{c.#1}}

\graphicspath{{figs/}}

\begin{document}

\title{
New Detectors to Explore the Lifetime Frontier
}

\author{John Paul Chou}
\email{johnpaul@physics.rutgers.edu}
\address{Department of Physics and Astronomy, Rutgers University, Piscataway, NJ 08854}

\author{David Curtin}
\email{dcurtin1@umd.edu}
\address{Maryland Center for Fundamental Physics, Department of Physics, University of Maryland, College Park, MD 20742 USA}

\author{H. J. Lubatti}
\email{lubatti@u.washington.edu}
\address{Department of Physics, University of Washington, Seattle, WA 98195}

\date{\today}
\begin{abstract}
Long-lived particles (LLPs) are a common feature in many beyond the Standard Model theories, including supersymmetry, and are generically produced in exotic Higgs decays. Unfortunately, no existing or proposed search strategy will be able to observe the decay of non-hadronic electrically neutral LLPs with masses above $\sim$ GeV and lifetimes near the limit set by Big Bang Nucleosynthesis (BBN), $c \tau \lesssim 10^7 - 10^8$~m. We propose the MATHUSLA surface detector concept (MAssive Timing Hodoscope for Ultra Stable neutraL pArticles), which can be implemented with existing technology and in time for the high luminosity LHC upgrade to find such ultra-long-lived particles (ULLPs), whether produced in exotic Higgs decays or more general production modes. We also advocate for a dedicated LLP detector at a future 100 TeV collider, where a modestly sized underground design can discover ULLPs with lifetimes at the BBN limit produced in sub-percent level exotic Higgs decays. 
\end{abstract}

\maketitle

\section{Introduction and Motivation}

The Standard Model (SM) of particle physics, completed by the 2012 discovery of a 125 GeV Higgs boson, is consistent with almost all phenomena observed at colliders in the last few decades.  In spite of this success, there are very strong reasons to suspect the existence of beyond-the-SM (BSM) physics, which could manifest at collider energies in many different ways. 
Long-lived particles (LLPs) with macroscopic decay lengths $\gtrsim$ $\mu$m are a particularly intriguing possibility, both due to their experimental distinctiveness and because they arise in BSM theories proposed to address virtually every fundamental mystery of particle physics.
This includes supersymmetric (SUSY) theories which address the well-known Hierarchy Problem, like mini-split SUSY \cite{Arvanitaki:2012ps, ArkaniHamed:2012gw}, gauge mediation \cite{Giudice:1998bp}, RPV SUSY \cite{Barbier:2004ez, Csaki:2013jza} and Stealth SUSY \cite{Fan:2011yu}. LLPs also feature prominently  in Neutral Naturalness  \cite{Burdman:2006tz, Cai:2008au, Chacko:2005pe}, Hidden Valleys~\cite{Strassler:2006im,Strassler:2006ri,Strassler:2006qa,Han:2007ae,Strassler:2008bv,Strassler:2008fv}, models of dark matter \cite{Baumgart:2009tn, Kaplan:2009ag,
  Chan:2011aa, Dienes:2011ja, Dienes:2012yz, Kim:2013ivd}, 
  explanations for the matter-antimatter asymmetry of the universe \cite{Bouquet:1986mq, Campbell:1990fa, Cui:2012jh, Barry:2013nva, Cui:2014twa, Ipek:2016bpf},  
  and models that generate neutrino masses \cite{Helo:2013esa, Antusch:2016vyf,Graesser:2007yj, Graesser:2007pc, Maiezza:2015lza, Batell:2016zod}.  

LLPs can be produced at colliders in many production modes. While electrically charged or hadronic LLPs leave a variety of signatures in the main detector, we focus on neutral\footnote{For our purposes, this general definition includes SM-singlets that arise e.g. in Hidden Sectors, but also particles which carry only $SU(2)_L$ charge like neutrinos, which can arise e.g. in SUSY.} LLPs, which can only be detected as displaced decays from the central interaction point, or as missing transverse energy (MET) if they escape. While searches for displaced decays are very challenging, significant progress has been made in recent years \cite{
ATLAS:2012av,
Aad:2012kw,
Aad:2012zx,
Aad:2015uaa, 
Aad:2015asa, 
Aad:2014yea, 
Aad:2015rba, 
Chatrchyan:2012sp,
Chatrchyan:2012jna,
CMS:2014wda, 
CMS:2014hka, 
Aaij:2014nma}, and experimental efforts at the Large Hadron Collider (LHC) and future colliders~\cite{Tang:2015qga, fcchhstudy, Burdman:2014zta, Dawson:2013bba} are likely to cover much of the possible signature space. 

Unfortunately, collider searches for LLPs face a fundamental limitation. A neutral LLP is a sterile state, completely non-interacting with SM matter and only visible once it decays. The decay length is an unknown parameter in many theoretical models. The only near-universal constraint comes from cosmology: Big Bang Nucleosynthesis (BBN), which occurred when the universe was about 1 second old, is very well understood within the SM and strongly constrained by measurements of cosmological parameters and elemental abundances~\cite{pdg}. 
LLPs decaying during or after BBN would change these observables. Avoiding this requires their lifetime to be in the range  $c \tau \lesssim 10^7 - 10^8$~m, unless their energy density is very small, their branching ratio to SM hadrons is very small, or they are cosmologically stable (see e.g. \cite{Kawasaki:2004qu, Jedamzik:2006xz}). 
Experimentally, it is encouraging that the range of allowed lifetimes is finite (up to the above caveats), but the ceiling of allowed decay lengths is frustratingly high. Therefore, ultra-long-lived particles (ULLPs) produced at the LHC will escape the main detector with extremely high probability. 

An escaping ULLP shows up as missing energy in the main detector.
If the ULLPs carry significant momentum and are produced with a high enough production cross section, LHC dark matter and/or SUSY searches would identify the ULLPs as characteristic MET.
However, in the event of such a discovery it would be unknown if the MET is due to an escaping stable particle or a ULLP, a distinction of great cosmological relevance. In the latter case, accurate characterization of the BSM sector would require direct detection of the ULLP decay.

In other scenarios, ULLP detection is even more urgent.  For example, relatively low-scale hidden sectors or squeezed spectra lead to ULLP production with momenta $\lesssim 100 \gev$, which generically makes MET searches at the LHC  extremely challenging. In this case, direct detection of the ULLP decay might be the only avenue to  their discovery.
 \emph{Exotic Higgs Decays} are perhaps the best motivated of these scenarios.\footnote{For very light LLPs like sub-GeV sterile neutrinos, fixed target experiments like SHiP \cite{Alekhin:2015byh} can be sensitive to BBN lifetimes.}

Searches for exotic decays of the recently discovered 125 GeV Higgs boson are one of the most promising discovery avenues for new physics~\cite{Curtin:2013fra}. The Higgs is copiously produced at $pp$ colliders through its coupling to the top quark via gluon fusion, but its decay width is dominated by the small bottom quark Yukawa $y_b \sim 0.02$. This means that even very small couplings of the Higgs to new degrees of freedom can manifest as exotic decay modes with observably large branching fractions. Such couplings are generic in BSM scenarios, since new hidden sectors that include neutral LLPs can couple to the Higgs via portal operators of low effective dimension.

In particular, this makes exotic Higgs decays one of the most plausible sources of LLPs. Furthermore, these LLPs will in general obey the BBN lifetime bound, since their coupling to the Higgs implies thermal contact with the SM plasma.\footnote{An exception to this argument may be inflationary scenarios with very low reheating temperatures.}
This is realized, for example, in theories of Neutral Naturalness \cite{Craig:2015pha, Curtin:2015fna, Csaki:2015fba,Chacko:2015fbc, Freytsis:2016dgf}, and generic scenarios with 
Hidden Valleys~\cite{Strassler:2006im,Strassler:2006ri,Strassler:2006qa,Han:2007ae,Strassler:2008bv,Strassler:2008fv} or  dark photons~\cite{Curtin:2014cca, Clarke:2015ala, Arguelles:2016ney}.
The large rate of Higgs and hence ULLP production can compensate to some extent for the low probability of decaying in the detector, 
but this is not enough to probe ULLPs with lifetimes near the BBN bound. 
This is especially regrettable, since invisible Higgs decays with $\sim 10\%$ branching ratios are discoverable at the High-Luminosity LHC (HL-LHC)~\cite{CMS:2013xfa,  ATL-PHYS-PUB-2013-014}. At future lepton colliders, sub-percent invisible decays can be directly probed~\cite{Dawson:2013bba, Gomez-Ceballos:2013zzn}.  For all of these MET signals, full characterization of the produced final state would be invaluable.

It is  for these reasons that we advocate in this paper for the construction of dedicated LLP detectors at hadron colliders. 
For the purpose of sensitivity estimates, we focus here on exotic Higgs decays as a benchmark model, where sensitivity to ULLPs at the BBN lifetime bound would allow for full coverage of an essentially finite parameter space. However, extending our sensitivity to long decay lengths is generally motivated to discover and diagnose many BSM scenarios.

At the HL-LHC, ULLP decays could be detected by instrumenting a suitably large volume near the interaction point, with sufficient shielding to suppress backgrounds from prompt particles produced in the collision.\footnote{This is similar in spirit to the MilliQan external detector proposed to find stable milli-charged particles produced at the LHC~\cite{Haas:2014dda, Ball:2016zrp}.} We propose the  \textbf{MATHUSLA} detector concept (MAssive Timing Hodoscope for Ultra Stable neutraL pArticles), which is located at the surface above and slightly displaced from the interaction point. This detector extends the lifetime range of LLP searches by three orders of magnitude compared to the main detector and could discover ULLPs near the BBN limit. With present technology and available land above the CMS or ATLAS interaction points, this detector could be constructed in time for the HL-LHC upgrade.

\section{MATHUSLA Surface Detector for the HL-LHC}

We propose constructing a dedicated detector to observe ULLPs decaying into charged SM particles away from the main detector. The HL-LHC will produce $N_h \approx 1.5 \times 10^{8}$ Higgs bosons. 
When producing ULLPs in exotic Higgs decays, the number of observed ULLP decays is roughly
\begin{equation}
\label{e.Nobs}
N_\mathrm{obs} \sim N_h \ \cdot \  \mathrm{Br}(h\to \mathrm{ULLP} \to \mathrm{SM}) \ \cdot \ \epsilon_{\mathrm{geometric}} \ \cdot \ \frac{L}{b c \tau}
\end{equation}
where $L$ is the linear size of the detector along the ULLP's direction of travel, $\epsilon_\mathrm{geometric}$ is the chance that the ULLP will pass through the detector (i.e. geometric coverage), and $b$ is the Lorentz boost $|\vec p|/m$ of the produced ULLP.
 Since the Higgs boson is dominantly produced on threshold, if it decays to $n$ ULLPs with mass $m_X$, their characteristic boost will be
\begin{equation}
\label{e.boost}
b \sim \frac{m_h}{n m_X},
\end{equation}
so typically $b \lesssim 3$ for $n = 2$ and $m_X \gtrsim 20 \gev$.\footnote{$n > 2$ typically implies higher-dimensional operators, cascade decays in the hidden sector, or a hidden confining gauge group that produces a shower. 
We focus on $n = 2$ which suffices to discuss detector requirements.} Assuming  the ULLP decays exclusively to the SM,\footnote{Generically, a coincidence of small widths is required for a ULLP to decay to both SM and hidden sector particles.} observation of a few ULLP decays with a lifetime of  $c \tau \sim 10^7$~m requires
\begin{equation}
 L \sim (20 \ \mathrm{m})  \left(\frac{b}{3}\right)  \left( \frac{0.1}{\epsilon_\mathrm{geometric}} \right) \frac{0.3}{\mathrm{Br}(h \to \mathrm{ULLP})} .
\end{equation}
Therefore, ULLPs near the BBN lifetime bound 
arising from exotic Higgs decays near current limits \cite{CMS-PAS-HIG-16-009, Aad:2015pla} 
could be discovered if the detector had a linear size of $\sim$20~m in the direction of travel and $\sim$10\% geometric coverage. 
Central to this estimate is the assumption that the dedicated ULLP detector operates in a very low background regime.
A necessary requirement is that the detector is shielded from the background of hadronic particles produced in association with the $pp$ collisions.

\begin{figure}
\begin{center}
\includegraphics[width=0.47 \textwidth]{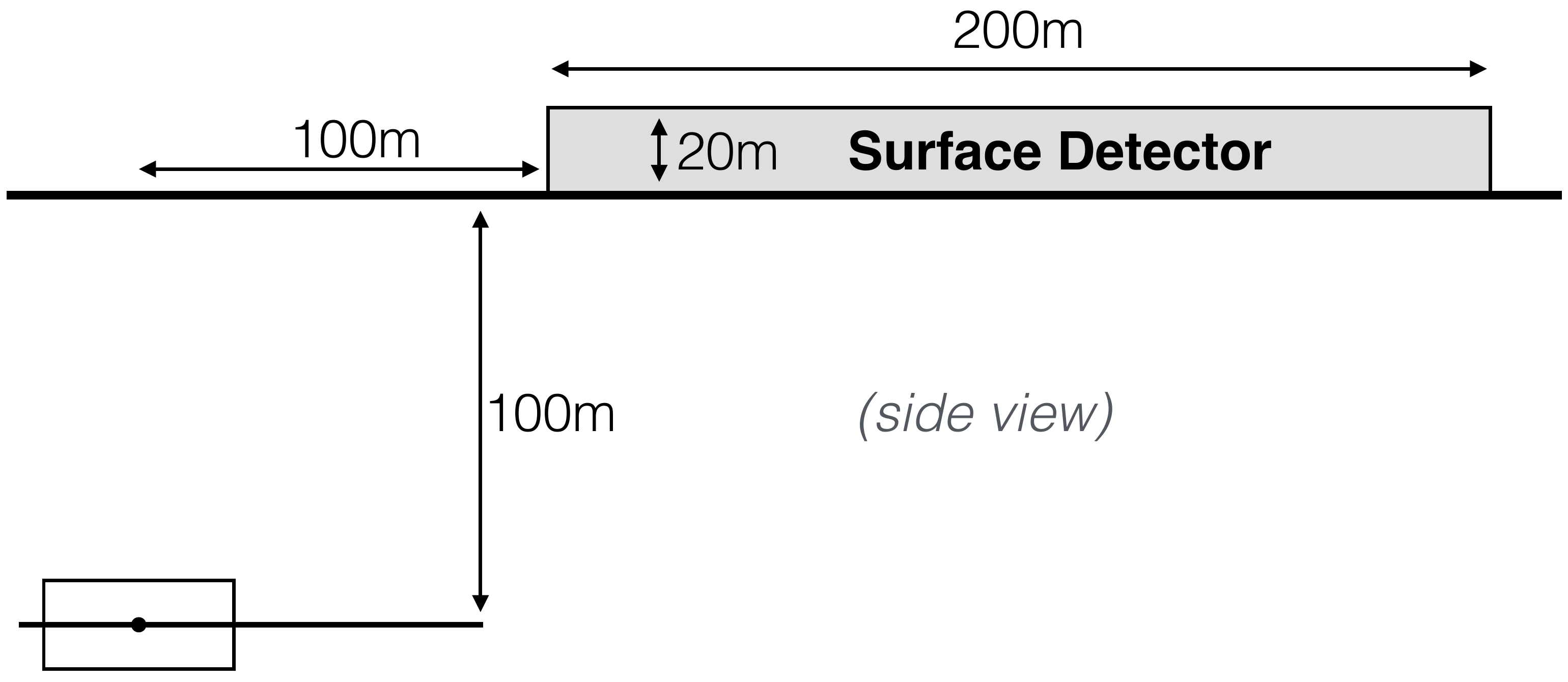}
\end{center}
\caption{
Possible geometric configurations for the MATHUSLA surface detector at the HL-LHC.  Gray shading indicates areas assumed to be sensitive to LLP decays. The surface detector is a 200m square building,  centered along the beam line.}
\label{f.detectorHLLHC}
\end{figure}

Placing the detector on the surface automatically satisfies the shielding requirement, but the $\mathcal{O}(100 \ \mathrm{m})$ distance from the interaction point requires a very large detector with an area of order $( 200\ \mathrm{m})^2$ and about $20$~m height to achieve a sufficiently large decay volume, as shown schematically in \fref{detectorHLLHC}. The precise location on the surface is not critical, and the volume could be broken up into submodules, provided that the horizontal distance is $\lesssim 100$ m. 
Consequently, for a given $\Delta \phi$ coverage, the rapidity of the detector does not greatly affect geometric acceptance to LLP decays.\footnote{For example, if LLPs are produced in exotic Higgs decays, simulations show that their rapidity distribution, weighted by inverse boost, is approximately uniform.} It may, however,  be convenient to center MATHUSLA near $\eta = 0$.

\begin{figure}
\begin{center}
\hspace{-0mm}
\includegraphics[width=0.48 \textwidth]{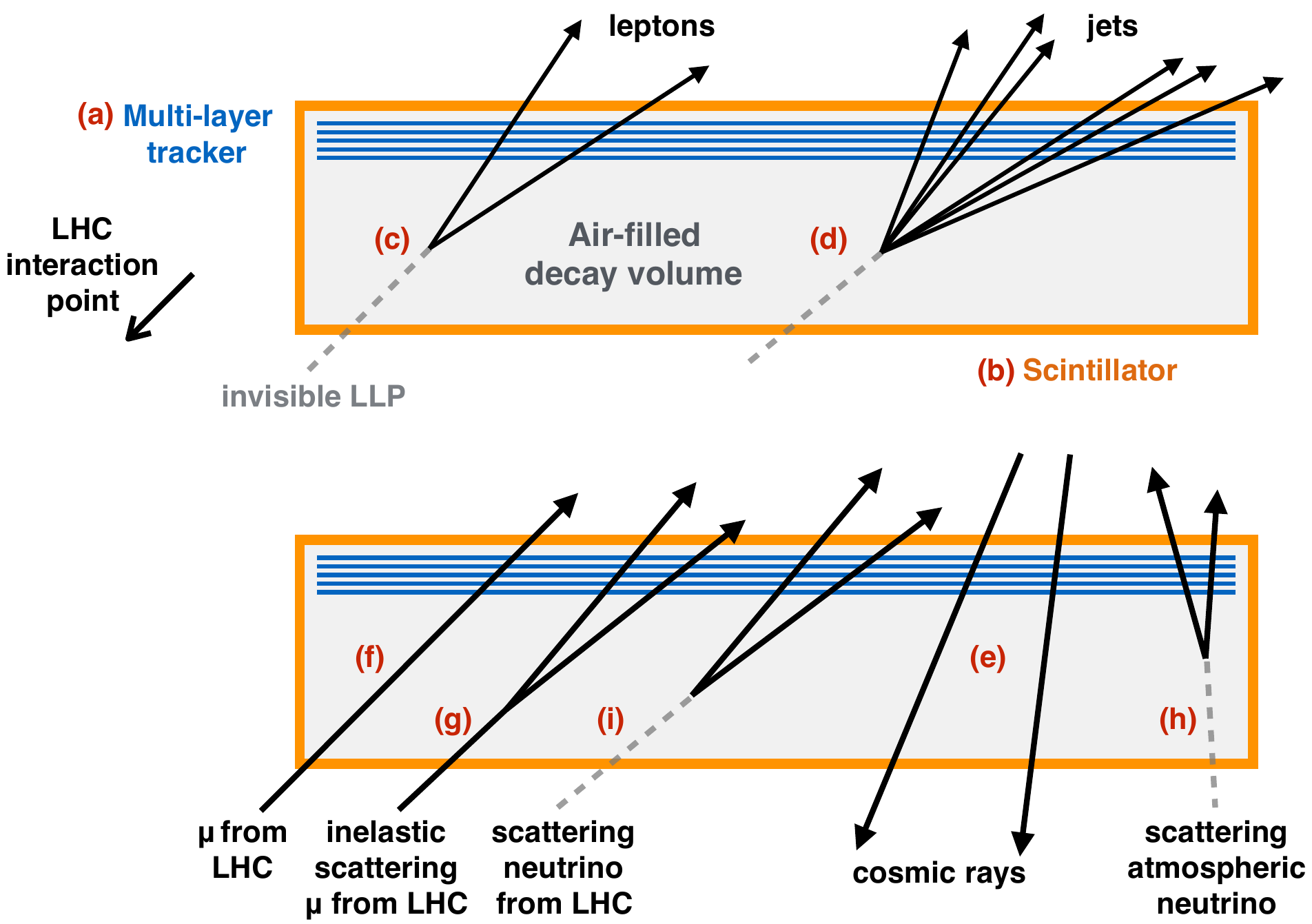}
\end{center}
\vspace*{-4mm}
\caption{
Schematic of a possible design for MATHUSLA, with a robust multi-layer tracker at the top and a segmented scintillator veto surrounding the entire detector. Also shown are two possible displaced vertex signals from LLP decays (top) and the five most important backgrounds (bottom). Black arrows indicate charged particles and their direction of travel.
}
\label{f.mathusla}
\end{figure}


The final design of the MATHUSLA detector and its associated signal acceptance and background rejection requires a detailed simulation study and small-scale test that is beyond the scope of this paper.
Here we outline a possible design that makes use of existing and well-tested technologies to demonstrate that the proposed approach will identify clearly LLP decays and provide the information needed to reject expected backgrounds.

\fref{mathusla} shows one possible instrumentation of MATHUSLA, assuming the geometry of \fref{detectorHLLHC}.
The air-filled-decay volume is surrounded by 1 cm thick plastic scintillator tiles (top, bottom and sides) with a robust, highly redundant, multilayer tracker located at the top. The tracking system allows for reconstruction of the trajectories of charged particles and provides timing information.  The separation between tracking planes can be of order 1 m or somewhat less.
One well-established candidate technology for this large area tracker are Resistive Plate Chambers (RPCs) that have cm spatial and ns timing resolution. RPCs have been deployed in detectors with areas greater than $\sim 5000~\mathrm{m}^2$~ \cite{Aielli:2006cj, Iuppa:2015hna}.

For now, we assume on the order of $\sim 5$ layers of tracking chambers, although 
the number in a final design may be different depending on the outcome of a detailed background rejection study. This combination of tracking and timing information from the tracking system and surrounding scintillator detectors provides information for rejecting (vetoing) backgrounds. 
For read-out, the scintillator detectors will be segmented, with the segment size requiring  detailed study. The tracking chamber readout geometry depends of the choice of technology, but in general terms this is unlikely to be a problem due to the low occupancy of the  detector.\footnote{Depending on the angle, a cosmic ray takes $\sim 70 - 100$ ns to traverse the detector volume. Therefore, given the $\mathcal{O}(10\  \mathrm{MHz})$ incident rate, the average cosmic ray occupancy rate of the entire detector in a 100 ns window is order one. 
A variety of segmentation strategies can easily accommodate the read-out requirements of such an event rate, even taking the much larger peak occupation during high-multiplicity shower events into account.}

There are many options regarding the construction of the detector structure. For example, partial burial may be preferable given its large volume. At any rate, the civil engineering requirements are modest.
Given that the proposed detector technologies are very well understood, MATHUSLA could feasibly be built in time for the HL-LHC upgrade.

Neutral LLPs decaying visibly in MATHUSLA result in a spectacular signal, with two examples  shown in \fref{mathusla} (c) and (d). A spray of charged particles, two in the case of lepton final states and $\mathcal{O}(10)$ in the case of hadronic final states, travel upwards due to the LLP's boost. 
The large area coverage and good tracking capabilities of MATHUSLA allow most of the charged particle trajectories to be individually reconstructed with detailed timing information. The trajectories can be fitted to reconstruct a displaced vertex (DV), with the additional stringent requirement that all trajectories are coincident \emph{in time at the DV}. 
The scintillator can be used as a veto to ensure that the charged particles \emph{originated} at the DV: there should be no hits along the line between the vertex and the LHC main interaction point (IP), as well as along the lines obtained by extrapolating the charged particle trajectories backwards. 
These are powerful signal requirements, especially for hadronic LLP decays.
The DV from an LLP decaying to leptons only has two charged particles associated with it, but it must satisfy additional geometric requirements: the line from the IP through the DV must lie in the plane spanned by and between the two charged particle trajectories.

By far the largest background is from atmospheric cosmic ray muons, see \fref{mathusla} (e). They have a rate of $\mathcal{O}(10\  \mathrm{MHz})$ over the entire detector~\cite{pdg}, resulting in  $\sim 10^{15}$ charged particle trajectories integrated over the whole the HL-LHC run. 
A very high rejection efficiency of this background is possible using a combination of timing and trajectory reconstruction because of the highly distinct signatures produced by cosmic ray muons compared to the decay of an LLP.

The vast majority of cosmic rays travel down, not up like final states from the decay of an LLP originating at the $pp$ IP. 
%
The chance that a \emph{single} downward-traveling cosmic ray will be mis-identified as originating within the detector volume and traveling upward is less than $10^{-15}$.
This rate was estimated by conservatively assuming that only three of the five RPC layers fire, with an assumed 1m spacing between layers and a timing resolution of 1 ns. 
We assume also that the scintillator layer at the bottom of the detector has an order percent chance of not detecting the passing charged particle.
 For this estimate, spatial information was not used at all: the signal requirement is only that the bottom scintillator layer is not activated, and that each RPC hit occurs within two timing resolutions of a consistent time for an upwards traveling particle with constant relativistic speed. This timing requirement, which could possibly be loosened by taking spatial information into account, does not significantly reduce signal efficiency at our level of precision. 

This small rate suggests that the only realistic way for cosmics to possibly fake the signal are correlated showers of many muons coincident on MATHUSLA. 
 However, these can easily be vetoed online and offline, due to their much higher occupancy compared to signal events, as well as the numerous stringent signal requirements outlined above.

The required very high efficiency rejection of cosmic ray backgrounds for LLP searches during the HL-LHC run will be a major driver of the detailed MATHUSLA design. The precise required timing and spatial resolutions of the tracker and scintillators, as well as the exact number of tracking layers, is the key question to be resolved by future detailed studies.\footnote{Adding a tracking layer at the bottom of the detector, for example, is an item to study.} 
However, the unique geometrical nature of the LLP DV signal, and the multitude of coincidences that would be required for cosmic rays to fake the signal, makes us confident that this background can be controlled. Furthermore, this background can be measured in detail in the absence of HL-LHC collisions. Such non-collision data also will aid in defining and/or validating rejection strategies.

Another important background is prompt high energy muons produced at the HL-LHC, shown in \fref{mathusla} (f). Using the known stopping power $\langle -dE/dx\rangle$ for muons in matter~\cite{pdg}, we estimate that a muon has to have an initial energy greater than $\sim 60 \gev$ to traverse more than 140 meters of rock and reach MATHUSLA. 
The SM production cross section for such high energy muons, dominantly from Drell-Yan processes, is readily estimated to lowest order in MadGraph 5 \cite{Alwall:2014hca}. Assuming an instantaneous luminosity of $5\times 10^{34} \mathrm{cm}^{-2} \mathrm{s}^{-1}$, this corresponds to a rate of $\mathcal{O}(10\ \mathrm{Hz})$. 
These muons do not reconstruct a displaced vertex. Therefore, by themselves, they do not satisfy the signal requirement.
However, the inelastic scattering of muons in the air could result in a reconstructed displaced vertex and rejection would rely solely on the scintillator veto (as well as possible timing or geometric cuts, see below).  However, the total number of these events during the HL-LHC run, $N_{\mu_\mathrm{scat}}$, can be estimated by treating MATHUSLA as a fixed-target experiment:
\begin{equation}
N_{\mu_\mathrm{scat}}\sim n_\mathrm{target} L_\mathrm{target} N_\mu \sigma_{\mu_\mathrm{scat}} ,
\end{equation}
where the scattering cross section $\sigma_{\mu_\mathrm{scat}}$ is in the $\mu$b range~\cite{Timashkov:2006kk}, $n_\mathrm{target}$ is the nuclear number density of air, $L_\mathrm{target} \sim 20 \mathrm{m}$ is the height of MATHUSLA, and $N_\mu$ is the total number of muons incident on MATHUSLA during the entire HL-LHC data taking period. The results is about $N_{\mu_\mathrm{scat}}\sim  10$ for the entire HL-LHC run, easily rejected by the scintillator detectors.

The MATHUSLA detector is immersed in a diffuse neutrino flux generated by cosmic ray interactions in the atmosphere~\cite{Daum:1994bf, GonzalezGarcia:2006ay,  Halzen:2013dva}. This flux is isotropic below PeV energies. Atmospheric neutrinos scattering off nuclei in the air-filled decay volume are a potentially dangerous background, since they can result in a topology similar to an LLP decay, see \fref{mathusla}~(h). A careful examination of these scattering events indicates that they can be distinguished from LLP decays, by the tracking and time-of-flight information.
Here, we estimate the rate of these neutrino scattering processes in MATHUSLA, as well some simple rejection strategies. We leave detailed analysis using dedicated neutrino Monte Carlo generators like GENIE~\cite{Andreopoulos:2015wxa} to a future study.
 
We restrict ourselves to neutrino energies greater than $\sim 400 \mev$, for reasons that will become clear below. In this region, the atmospheric muon neutrino flux as measured by Frejus~\cite{Daum:1994bf, Halzen:2013dva} is reasonably well approximated by
\begin{equation}
\frac{d \Phi}{d E_\nu} \sim 0.06 \left( \frac{\gev}{E_\nu}\right)^3 \ \gev^{-1} \mathrm{cm}^{-2} \mathrm{s}^{-1} \mathrm{sr}^{-1}
\end{equation}
(though we use the actual measured neutrino flux spectrum in our calculations). The electron neutrino flux is significantly lower and can be treated similarly, so we neglect it in this simple estimate.
For a given neutrino scattering process $i$ with cross section $\sigma_i$, the number of produced events \emph{per year} can be estimated as
\begin{equation}
N_i =  \int d E_\nu \  \frac{d \sigma_i}{d E_\nu} \ n_\mathrm{target}\  \left[ 4 \pi \frac{d \Phi}{d E_\nu}  \ V_\mathrm{target} \ T\right ]
\end{equation}
where $V_\mathrm{target}$ is the volume of the MATHUSLA detector as per \fref{detectorHLLHC}, the exposure time $T$ corresponds to the fraction of the year the HL-LHC beams are running (assumed to be 50\%), and $n_\mathrm{target}$ is the number density of protons and neutrons in air. The various differential neutrino scattering cross sections $d\sigma_i/d E_\nu$ are theoretically and experimentally known at the better than $\sim 30\%$ level and are summarized in \cite{Formaggio:2013kya}. In practice we distinguish between protons and neutrinos, but we do not account for the fact that these are bound in nuclei. (As we note below, this underestimates our ability to reject these backgrounds.) We are interested in processes $i$ with at least two charged particles in the final state, including deep-inelastic scattering (DIS), so they can reconstruct a DV. We then divide those processes into two classes: those which are exclusively defined to contain a proton in the final state (PFS), like many quasi-elastic scattering (QES) processes, and those which are not (like DIS).

The number of scatters per year with protons in the final state is 
\begin{equation}
\sum_{\mathrm{PFS}} N_i \approx 60 \ ,
\end{equation}
which is  non-negligible. 
In most of these scatterings the final-state proton is measurably non-relativistic, and the final states are contained within a very narrow cone that generally does not point back to the LHC IP. 
A veto of DVs with a narrow final state cone that does not point back to the IP has minimal effect on signal acceptances: light and highly boosted LLPs give a cone which points back at the IP, while somewhat heavier LLPs give final states which are not confined to a single narrow cone. 
To make this more quantitative, assume these processes obey the differential cross section of QES.
%
We also assume, very conservatively, that a proton moving slower than $v_\mathrm{min} = 0.6c$ can be reliably distinguished from an ultra-relativistic muon. (This corresponds to a 10 ns difference in travel time over 5 meters, which is a reasonable estimate for the depth of the tracker at the top.) Only neutrinos with energy above about 500 MeV can produce such protons (hence we only considered that part of the neutrino spectrum). For protons faster than $v_\mathrm{min}$, $2\to2$ kinematics dictates a maximum cone size for the final states, with a corresponding small probability that this cone's random orientation points back at the LHC IP. The number of scatters per year that passes these cuts and could actually fake a DV signal is
\begin{equation}
\label{e.Naftercuts}
\sum_{ \mathrm{PFS}} N_i^\mathrm{after\ cuts} \approx 1  \ .
\end{equation}
It is also important to point out that this low rate is a significant \emph{overestimate}, since (1)~our assumption that MATHUSLA could only distinguish $0.6c$  from $c$ is very conservative; (2) for non-QES processes the additional produced particle will result in slower final-state protons and/or narrower final-state cones; (3) we assumed that all scatters with protons faster than $v_\mathrm{min}$ had the largest possible final-state cone assuming $v_\mathrm{proton} = v_\mathrm{min}$; and (4) we neglected that the initial-state protons and neutrons are bound in nuclei, and the  slow-moving debris from the breakup of the nucleus could provide additional rejection handles.

We now turn to atmospheric neutrino scatters without a definite proton in the final state, for which the rate per year is
\begin{equation}
\sum_{\mathrm{not \ PFS}} N_i \approx 10 \ .
\end{equation}
It is difficult to make detailed statements about these events without simulations, but they are dominated by DIS with neutrino energies in the several GeV range. We therefore expect that the final states will be confined in a relatively narrow cone, which only rarely points back to the LHC IP. 
Obviously, more detailed studies are required to make these predictions  robust, but from these estimates we expect that of all the different kinds of atmospheric neutrino scattering events in MATHUSLA can be rejected, with none or very few events surviving as irreducible backgrounds over the entire HL-LHC run. 
Furthermore, it is important to recall that this background, like charged cosmic rays, can be measured in detail when there are no colliding beams in the HL-LHC.

Cosmic rays are not the only source of neutrinos that are incident on MATHUSLA. They are also produced at the HL-LHC, see  \fref{mathusla}~(i).
Direct production of neutrinos in $W$, $Z$, top and bottom quark decays can be estimated in MadGraph 5. The total number of these relatively high-energy neutrinos which scatter off air in MATHUSLA, computed in a similar manner as for  muons and atmospheric neutrinos above, is less than $0.1$ events over the entire HL-LHC lifetime. 
In addition, there is a low-energy neutrino flux ($E_\nu \lesssim 1 \gev$) from decay of pions and kaons produced in primary and secondary hadron production in the LHC.  We estimate this rate by simulating minimum-bias collisions with Pythia 8~\cite{Sjostrand:2007gs} and propagating the particles through the approximately azimuthally symmetric material budget~\cite{Bayatian:2006zz} of the CMS experiment with GEANT4~\cite{Agostinelli:2002hh}.  
The vast majority of LHC neutrinos are produced by low-energy hadrons decaying in the LHC detectors.
This leads to a steeply falling neutrino spectrum for energies above a few hundred MeV.  These neutrinos have mainly quasi-elastic neutrino-nucleon scatterings in MATHUSLA. Imposing the same proton-time-of-flight cut used to derive \eref{Naftercuts} (but not imposing any geometrical cuts since the final state cone always points back to the IP), we arrive at an (over-) estimate of $\approx 1$ scattering event with protons in the final state passing cuts \emph{over the entire HL-LHC run}. There are also $\approx 3$ scattering events (no cuts) from processes like DIS, which mostly do not have protons in the final state. As discussed above, their rejection is harder to estimate, but given these very low rates, we are confident that rejection can be further improved with detailed study.

Lepton interactions in the rock below MATHUSLA could produce long-lived $K_L^0$ hadrons that can migrate into the detector and decay. 
The rate of these processes can be estimated using the cross sections in \cite{Formaggio:2013kya}, and is much lower than other scattering processes considered above. 
In addition, a significant number of the $K_L^0$ will interact in the rock and not make it to the surface.  Finally, the long-lived hadron is produced in association with other charged particles. If any of them make it into the detector along with the $K_L^0$ the scintillator veto would be triggered. We conclude that $K_L^0$ production by inelastic lepton interactions is not problematic.  Further detailed studies will quantify such rare processes. 

Electronics noise and other ambient, low-energy backgrounds will ultimately be addressed by balancing the costs of redundancy against higher performance detectors and electronics.

\begin{figure}[t!]
\begin{center}
\begin{tabular}{c}
\hspace{-3mm} \includegraphics[width=8.5cm]{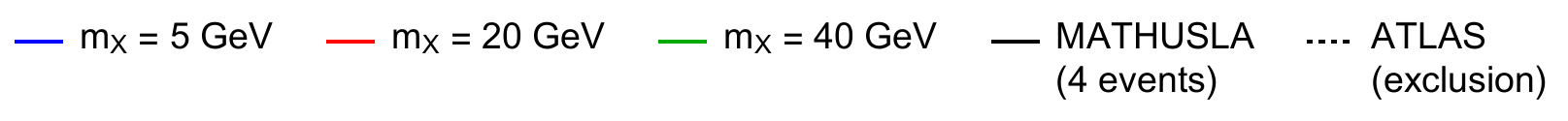}
\\
\includegraphics[width= 0.47 \textwidth]{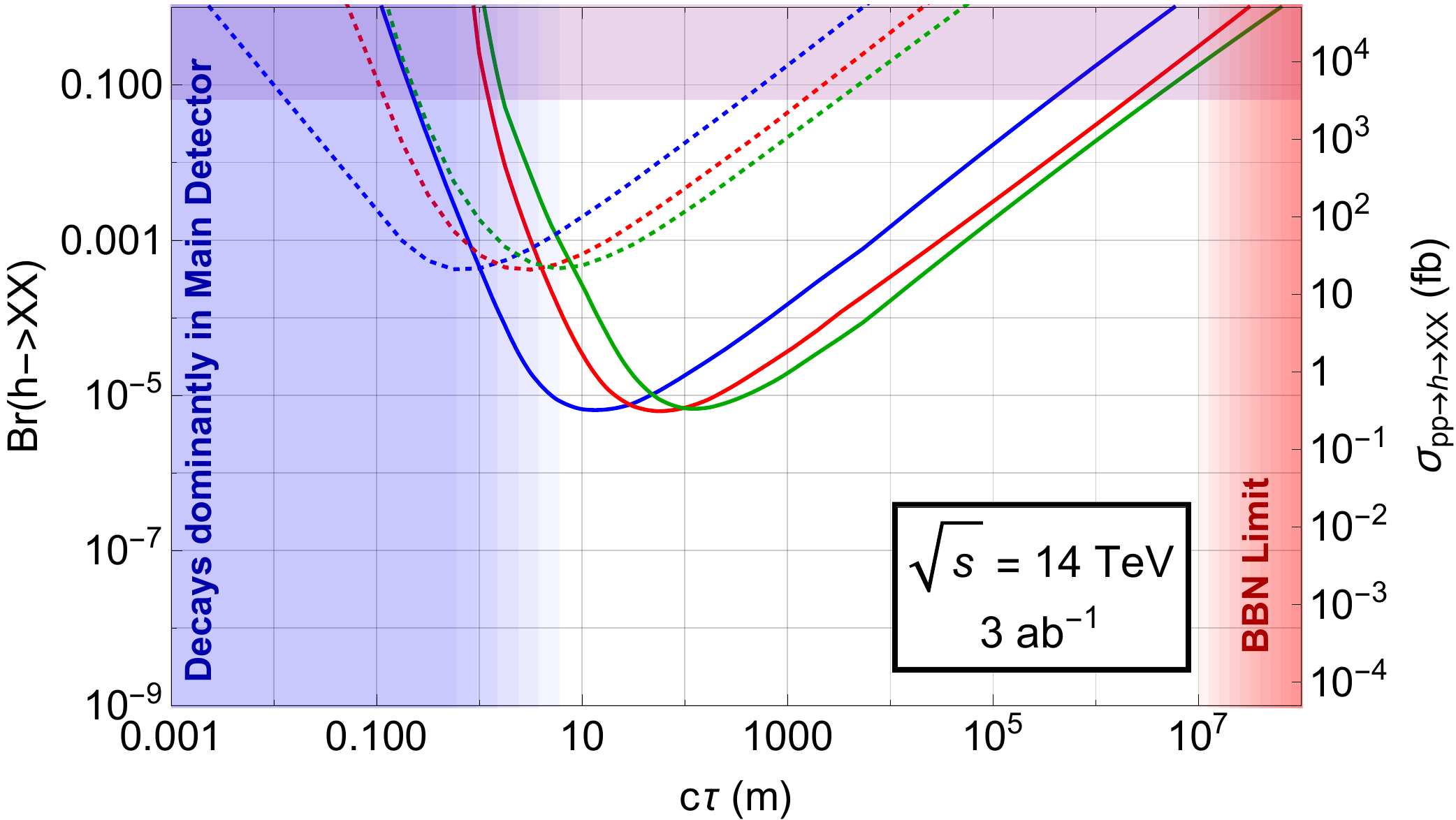}
\end{tabular}
\end{center}
\vspace*{-4mm}
\caption{
HL-HLC sensitivity to LLP production in exotic Higgs decays. 
Solid lines: Required $\mathrm{Br}(h \to XX)$ required to see 4 events in MATHUSLA. Dotted lines: projected ATLAS $\mathrm{Br}(h \to X X)$ exclusions~\cite{Coccaro:2016lnz}. 
Purple shading: projected CMS $\mathrm{Br}(h \to \mathrm{invis})$ exclusion~\cite{CMS:2013xfa}, which applies roughly beyond the blue shaded region.
}
\label{f.limitsHLLHC}
\end{figure}

We now present quantitative sensitivity estimates for the MATHUSLA detector design at the HL-LHC. 
To model LLP production in exotic Higgs decays, we assume the $h \to XX$  signal model, which can be applied to many more complete theories and demonstrates the reach of MATHUSLA in a simple parameter space. 
We simulate Higgs production via gluon fusion with subsequent decay to two LLPs $X X$ using the Hidden Abelian Higgs model \cite{Curtin:2014cca} in MadGraph 5 \cite{Alwall:2014hca}, matched up to one extra jet, and showered in Pythia 6~\cite{Sjostrand:2006za}. The probability of an LLP decaying in the MATHUSLA volume for a given lifetime is then straightforwardly computed by a convolution of the LLP kinematic distributions with the acceptance of the  geometry shown in \fref{detectorHLLHC}. 
The exact detection efficiency for LLP decays will depend on the final detector design, but we have checked that two or more final states of the decay hit the top of the detector $\sim 50\% - 95 \%$ of the time, depending on the LLP mass. 
The signal reconstruction efficiency could be very close to this, given the extremely robust tracking required for background rejection, and possibly even better  depending on the detailed design and geometry. 
Therefore, we assume an aspirational 100\% detection efficiency for simplicity, which suffices to demonstrate the achievable reach.

The $\mathrm{Br}(h \to XX)$ required for 4 expected signal events is shown as solid lines in \fref{limitsHLLHC} for different LLP masses $m_X = 5, 20, 40 \gev$.\footnote{Based on the required energy deposition and final state kinematics compared to the neutrino scattering background, we expect MATHUSLA to be sensitive to LLP masses above a few GeV, with sensitivity to lower masses possible in some cases. The details depend on the final detector design.} 
If LLP decays were actually observed, this would be within a modest $\mathcal{O}(1)$ factor of the discovery reach, assuming backgrounds could be almost completely rejected as we argue above. Conversely, if no events are observed, these curves correspond to 95\% CL exclusions, again assuming no backgrounds.

An important point of comparison is provided by the recent study \cite{Coccaro:2016lnz}, which investigated the sensitivity of a search for a single displaced vertex (DV) in the ATLAS Muon System (MS) with data-driven background estimates. Out of all the possible searches in the main detector, this strategy will have the best or close to the best sensitivity for long LLP lifetimes.
The projected exclusion limits from this search, making optimistic assumptions about QCD background, are shown as  dotted lines in \fref{limitsHLLHC}. The MATHUSLA detector extends lifetime sensitivity by  three orders of magnitude, and allows detection of ULLPs close to the BBN lifetime limit.

For decay lengths less than $\sim 10^{6}$ m, MATHUSLA is also orders of magnitude more sensitive than invisible Higgs decay searches. For relatively large $\mathrm{Br}(h \to \mathrm{ULLPs}) \gtrsim 10\%$, both searches would find a signal. Assuming the invisible Higgs decay products are all ULLPs, this would make it possible to uniquely determine the lifetime and  diagnose the hidden sector dynamics.

Our estimates demonstrate that a concept like MATHUSLA is capable of searching for and discovering LLPs with lifetimes approaching the Big Bank Nucleosynthesis limits.

\section{Sub-surface Detectors for the 100~TeV Collider}

 A 100 TeV collider with $30 \iab$ of integrated luminosity, such as the proposed SPPC \cite{Tang:2015qga} in China or the FCC-hh at CERN~\cite{fcchhstudy}, will produce  $N_h \approx 2.2 \times 10^{10}$ Higgs bosons. Correspondingly, the achievable sensitivity to ULLP decays is orders of magnitude greater than at the HL-LHC.  
At the BBN limit, the sensitivity to sub-percent level exotic Higgs decay branching fractions to ULLPs matches or exceeds invisible decay searches. For intermediate lifetimes, a dedicated ULLP detector could outperform invisible Higgs searches at lepton colliders by orders of magnitude. 
 
A surface detector like MATHUSLA could be constructed at a 100 TeV collider as well, but the required size may not be practical if the tunnel is deeper underground than the LHC. 
However, we argue that a dedicated \emph{underground} ULLP detector should be included in the design of any new $pp$ machine \emph{ab initio}, since the excavation of a new tunnel provides ample opportunity to repurpose an underground hall near the main detector for this purpose at relatively little additional cost. 
Such a subsurface design can be much more compact than MATHUSLA while achieving much greater LLP acceptance.

A benchmark geometry could be a Forward LLP detector reminiscent of a Muon System of HCAL endcap. For concreteness, we assume a detector volume in the shape of a cylindrical ring aligned with the beamline, with an inner radius of 5 m, an outer radius of 30 m, a depth of 20 m, and situated starting 40 m from the interaction point down the beamline. This should allow for at least 20 m of shielding, based on typical main detector geometries proposed for the 100 TeV collider~\cite{fcchhdetectorslides}.

Leaving more detailed design and background studies for future work, we make just a few remarks. Backgrounds from charged cosmic rays and atmospheric neutrinos are greatly reduced, due to the lower detector volume and, for the former, shielding by the rock above. A dominant background is likely to be muons produced in the 100 TeV collision, and highly redundant vetoes would have to be implemented accordingly. However, it is clear that in principle, similar detector technologies as those used in MATHUSLA should be sufficient to reject all backgrounds. Furthermore, the compact size of this detector may allow for more elaborate instrumentation, which would aid in the analysis of LLP decays as well as the rejection of backgrounds. 

We quantitatively estimate sensitivity in the same fashion as for the HL-LHC, with results for a MATHUSLA-like surface detector (assuming an area scaled up by the greater depth of the 100 TeV tunnel), and the benchmark Forward Detector, shown in \fref{limits100tev}. 
The projections in \fref{limits100tev} make clear that a dedicated ULLP detector at a 100 TeV collider could probe many ULLP production scenarios with lifetimes up to the BBN limit, including sub-percent exotic Higgs decays.

\section{Conclusion} 

LLPs occur in a wide variety of  BSM scenarios addressing the most fundamental mysteries of particle physics. Generally, their lifetime is only constrained by BBN, consequently for most of the allowable range the ULLPs escape the LHC detectors and only show up as MET. Direct ULLP detection is strongly motivated. In scenarios where MET searches are efficient, it is required to determine whether the escaping particle is unstable, or a dark matter candidate. 
In other well-motivated scenarios, including low-scale hidden sectors and squeezed spectra, MET searches are inefficient and direct ULLP detection may be the only discovery avenue for new physics. A well-motivated example is exotic Higgs decays, for which sensitivity to ULLPs near the BBN lifetime bound is possible. This allows for an essentially finite parameter space of new physics scenarios to be comprehensively covered.

We propose the MATHUSLA surface detector for the HL-LHC. Initial estimates suggest that it could search for LLPs in an extremely low background regime, and further work is underway to verify those claims. MATHUSLA would extend the sensitivity to long decay lifetimes by orders of magnitude compared to LHC detector searches, and could detect ULLP production cross sections in the $\sim$ 10~pb range for lifetimes near the BBN limit.
Even greater sensitivity is possible with a sub-surface design at a future 100 TeV collider.

\begin{figure}[t!]
\begin{center}
\begin{tabular}{c}
\hspace{-3mm} \includegraphics[width=8.5cm]{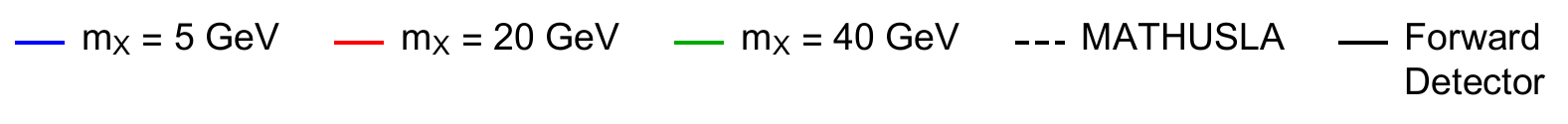}
\\
\includegraphics[width=0.47\textwidth]{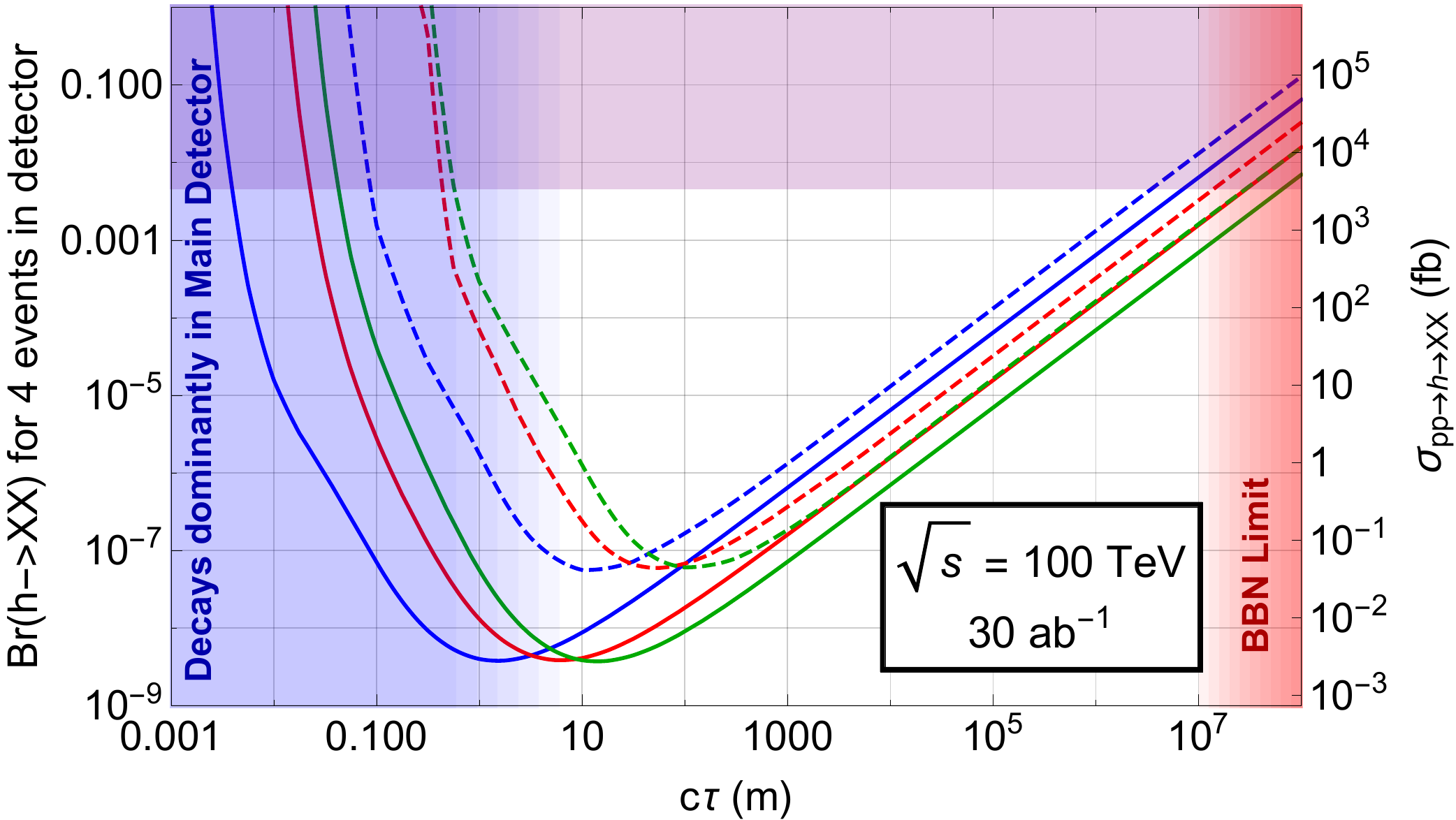}
\end{tabular}
\end{center}
\vspace*{-4mm}
\caption{
Projected 100 TeV reach for $\mathrm{Br}(h \to X X)$, the exotic Higgs branching fraction to LLPs. 
Purple shading represents the $\mathrm{Br}(h \to \mathrm{invis})$ bound projected for a TLEP-like future lepton collider \cite{Gomez-Ceballos:2013zzn}, which applies roughly beyond the blue shaded region.}
\label{f.limits100tev}
\vspace*{-3mm}
\end{figure}

\section*{Acknowledgements}
We thank
Nima Arkani-Hamed,
Zackaria Chacko,
Timothy Cohen,
Keith Dienes,
Rouven Essig,
Roni Harnik,
Christopher Hill,
Kara Hoffman,
Can Kilic,
Michelangelo Mangano,
Adam Martin,
Patrick Meade,
Prashant Saraswat,
Sunil Somalwar,
Matthew Strassler,
Raman Sundrum,
Scott Thomas,
and
Yuhsin Tsai
for useful conversation. 
We are very grateful to 
Timothy Cohen,
Keith Dienes,
Patrick Meade,
Aleandro Nisati,
Prashant Saraswat,
Andris Skuja,
and
Charlie Young
for helpful comments on a draft version of this letter. 
We also thank the Maryland Center for Fundamental Physics and the participants of the 2016 UMD Hidden Naturalness Workshop, where this collaboration was formed. 
JPC thanks Jim Freeman, Anna Pla-Dalmau, and Jim Hirschauer for help in obtaining cost estimates for large volumes of plastic scintillator, photosensors, and readout electronics.
DC thanks Ivan Poloni for helpful conversation on civil engineering aspects of the detector construction. 
HL thanks Rinaldo Santonico and Ricardo Cardarelli for providing cost estimates for a fully instrumented, large-volume RPC detector and informative discussions on the intrinsic spatial resolution of RPCs.
HL and JPC thank colleagues in the ATLAS and CMS Exotic Groups for useful discussions. 
DC and JPC thank KITP at the University of California Santa Barbara, supported in part by the National Science Foundation under Grant No. NSF PHY11-25915, where part of this work was completed. 
DC also thanks the Aspen Center for Physics, which is supported by National Science Foundation grant PHY-1066293, where part of this work was completed.
The work of JPC is supported by the NSF under grant No. PHY-1607096 and the Alfred P. Sloan Research Fellowship FG-BR2014-041.
D.C. is supported by National Science Foundation grant No. PHY-1315155 and the Maryland Center for Fundamental Physics.  
H.L. is supported by  National Science Foundation grant PHY-1509257.

\bibliography{lifetimefrontier}

\end{document}